# User-independent optical path length compensation scheme with sub-ns timing resolution for 1 × N quantum key distribution network system

**Byung Kwon Park,[1,2] Min Ki Woo,[1] Yong-Su Kim,[1,2] Young-Wook Cho,[1] Sung Moon,[1] Sang-Wook Han[1,*]**

[1]*Center for Quantum Information, Korea Institute of Science and Technology, Seoul 02792, South Korea*
[2]*Division of Nano and Information Technology, Korea Institute of Science and Technology School, Korea University of Science and Technology, Seoul 02792, South Korea*
**Corresponding author: swhan@kist.re.kr*

**Quantum key distribution (QKD) networks constitute promising solutions for secure communication. Beyond conventional point-to-point QKD, we developed 1×N QKD network systems with a sub-ns resolution optical path length compensation scheme. With a practical plug-and-play QKD architecture and compact timing control modules based on a field programmable gate array (FPGA), we achieved long term stable operation of a 1×64 QKD network system. Using this architecture, 64 users can simultaneously share secret keys with one server, without using complex software algorithms and expensive hardware. We demonstrated the working of a 1×4 QKD network system using the fiber network of a metropolitan area.**

## 1. INTRODUCTION

Quantum key distribution (QKD) networks can be promising solutions for secure communication, even in homes [1–6]. For ensuring secure communication with several distinct parties, various networks have been proposed, including field test QKD networks [7–11], free-space QKD networks using satellites [12], more secure QKD networks to defeat quantum hackings [13], and QKD network architectures combining classical communication schemes [14–16]. As a result of several experimental and theoretical efforts, QKD network technology is now considered to be ready for commercialization.

There are some reported QKD network schemes that utilize beam splitters, which are adoptable with simply a passive component between a sender and a receiver [5,17,18]. However, the more channels that exist, the more serious the problem, because the signal loss is proportional to the number of participants in the network. To overcome this, optical active switches have been applied [19,20]. For faster communication and switching, network schemes with wavelength division multiplexing (WDM) have been studied [21–24]. Optical ring architectures have also been reported [25–27]. These network architectures can be categorized into two types: One is based on point-to-point connections, called trusted nodes, and the other involves one-to-many or many-to-many network architectures. The network architecture built using point-to-point QKD has a high key rate and exhibits stable operation; however, it is expensive and bulky, making it difficult to establish the several linkages required between the trusted nodes. In contrast, the one-to-many or many-to-many network architecture is cost efficient and easily scalable; however, there are some technical difficulties caused by increasing number of users, such as problems regarding key generation speeds, crosstalk noise, and channel capacity. The one-to-many network architecture is useful in massive quantum networks for linking large numbers of end-users, such as those in data centers and military communication. The servers of one-to-many network architectures should have the ability to stably manage a large number of users.

Compared with one-way architectures, the plug-and-play quantum key distribution (P&P QKD) has several advantages when using the one-to-many quantum network architecture [28,29]. First, the server easily manages all users with an optical path length compensation algorithm for detecting the photons from the users, because a light signal is generated and detected in the server [30]. In contrast, one-way QKD has to consider the photon-arrival time, phase fluctuation, and polarization of photons from each user; these are important issues when there are several users on the network. In addition, multiple users do not need expensive hardware, such as avalanche photodiodes (APDs), and a light source is required only in a single server. The optical elements for the users comprise only some passive components, and several active components, namely phase modulator (PM) and intensity modulator (IM), for preparing quantum states and decoy pulses [31–34], respectively. Therefore, the optical elements and control units for the users can be built on a small scale.

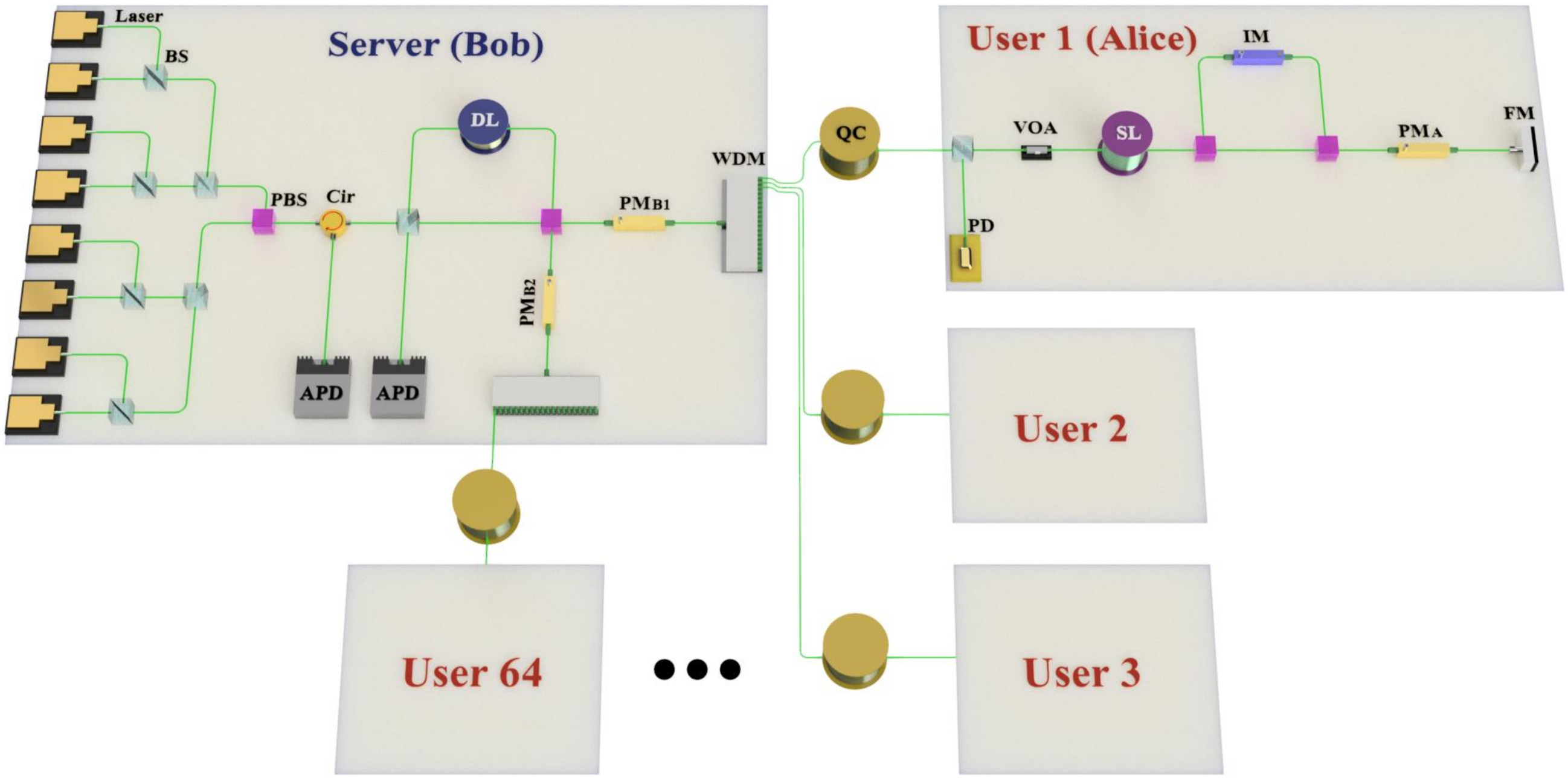

Fig. 1. P&P QKD network system architecture for 64 users. The coherent light signals from the server, Bob, are transmitted to 64 users, including Alice, using WDM and polarization division multiplexing. All users have identical optical systems. BS: beam splitter, PBS: polarization beam splitter, Cir: circulator, APD: avalanche photodiode, DL: delay line, PM: phase modulator, WDM: wavelength division multiplexing, QC: quantum channel, PD: photodiode, VOA: variable optical attenuator, SL: storage line, IM: intensity modulator, and FM: Faraday mirror.

In this study, we developed a 1 × N QKD network based on a P&P QKD architecture. For managing the arrival time of the photons from the users, we applied compact optical path length compensation modules based on an FPGA. The system performances of each user, such as the sifted key rate and quantum bit error rate (QBER), were monitored to compensate fluctuations in the optical path length due to environmental changes. Moreover, we demonstrated the effectiveness of this system in an actual network over a week, while maintaining stable system performance.

The remainder of this paper is organized as follows: In section II, we explain the overall architecture of the proposed QKD network system. In section III, we discuss the experimental and practical application results. Finally, in section IV, we summarize and conclude our work.

## 2. OVERALL ARCHITECTURE

Fig. 1 depicts the developed QKD network system architecture. To realize the 1 × N QKD network, we used WDM and polarization division multiplexing (PDM) together to support up to 64 users. In our setup, the PDM duplicates channel capacity, in contrast to using only WDM. The server, Bob, has eight tunable lasers to transmit light to Alice, the user. Each tunable laser can emit week coherent light signals with eight different wavelengths corresponding to eight users. All 64 light signals from the eight tunable lasers are controlled via a temperature control module in the laser. These signals from the tunable lasers can be enabled or disabled independently. They are merged using six beam splitters (BSs). Thirty two week coherent light signals from four of the tunable lasers have vertical polarization, and the others have horizontal polarization, achieved using the first polarization beam splitter (PBS). After traversing a circulator and an interferometer, at the second PBS, the lights with 32 different wavelengths are transmitted to the WDM on the right or down sides, according to the polarization of the light signals. To avoid crosstalk with adjacent channels, we rigorously chose 32 wavelength channels through temperature control of the tunable lasers. The lights from four tunable lasers were distributed to 32 users, through the right- or down-side WDM. Upon reaching Alice over the quantum channel (QC), the light is split by the BS to detect the synchronization signal emitted by the photodiode (PD). The residual light is kept in the storage line (SL) and passes the two PBSs and intensity modulator (IM) to generate the decoy pulse. We used two PBSs and an IM because of the polarization dependence of the IM. The light is encoded and reflected at the phase modulator ($PM_A$) and Faraday mirror (FM), respectively. At the variable optical attenuator (VOA), the light is attenuated to a single photon level and transmitted to Bob. At Bob's side, the photons arriving from 64 users are modulated at the right ($PM_{B1}$)- or down ($PM_{B2}$)-side PM and interfered at the BS of the interferometer. The arriving photons are detected by a pair of APDs with time division multiplexing (TDM) to enable low cost implementation.

To detect the photons from 64 users with a single pair of APDs, the time scheduling technique was adopted. The server recognizes the roundtrip time of the photons using the distance of each user. Therefore, the server generates laser pulses at the appropriate time and detects the photons the using gate pulse that is generated at a predetermined time slot. This is important for the stable operation of QKD network systems. To control the generation timing of the laser pulse, we developed 64 timing control modules using an FPGA, as shown in Fig. 2. These modules operate independently to control the laser timings for each user. The serializer in the timing control module can make the server arbitrarily generate high frequency signals. The serializer comprises a phase locked loop (PLL), D-flipflop (D-FF), and transmitter. A data uploader module transfers parallel data to the serializer for generating high speed serialized data at approximately 1.6 GHz. The server manages operation scheduling for users by allocating laser pulses at appropriate times with the serializer. Moreover, the laser frequency can be adjusted using the serializer, to optimize the user capacity of the TDM. For example, if four users need to distribute quantum keys, the server operates each laser at 1/4 of the maximum system frequency. In addition, as mentioned before, environmental changes around optical fibers result in incorrect laser timing owing to variations in the effective optical path length. In our previous work, we compensated the variation in the optical length using only a serializer [30]. To enhance the resolution of the serializer, we serially added functions with delay chains based on the FPGA. The serializer outputs high speed low voltage differential signals (LVDSs), which are fed again to the FPGA, owing to the incompatibility of the serializer with the delay chain. The delay chain consists of IO configuration, dynamic delay, and output buffer (Obuf) blocks. The delay control block communicates with the IO configuration block to control two dynamic delay blocks. The dynamic delay blocks are serially connected to delay the signals up to 22 steps with a 50 ps resolution for one step; this can fully cover the minimum resolution of the serializer. This means that the signal can be generated at an overall timing with a 50 ps resolution. Using the functions of the serializer and delay chain, the server not only independently controls the laser pulses for all users, but also tracks variations in the optical length with a 50 ps resolution without any additional hardware. The output signals from the delay chain are fed to the amplifier (AMP), which generates well maintained signals for operating the lasers.

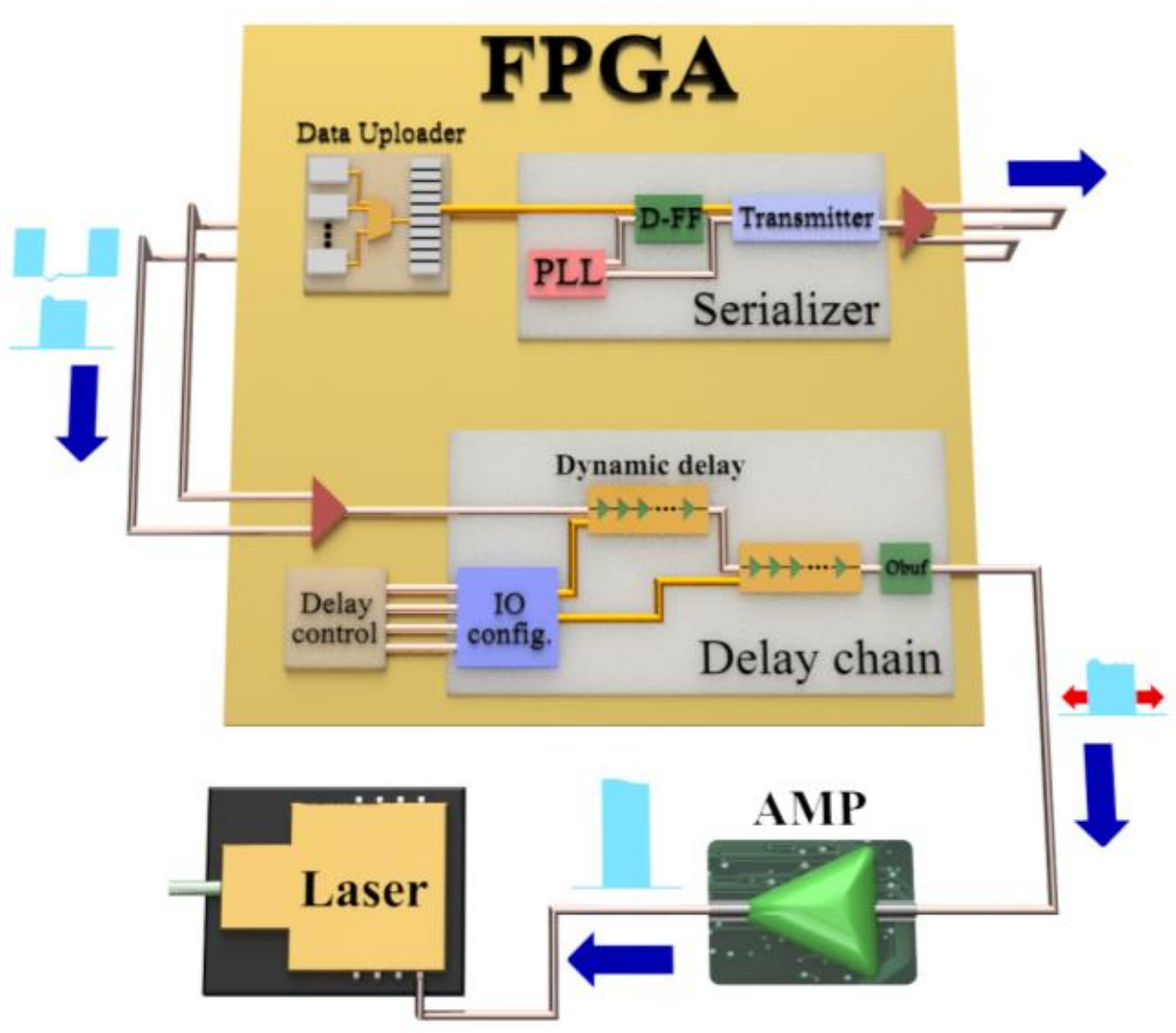

Fig. 2. Timing control module based on an FPGA for one user. The arrows indicate the flow of the signal for generating laser pulses.

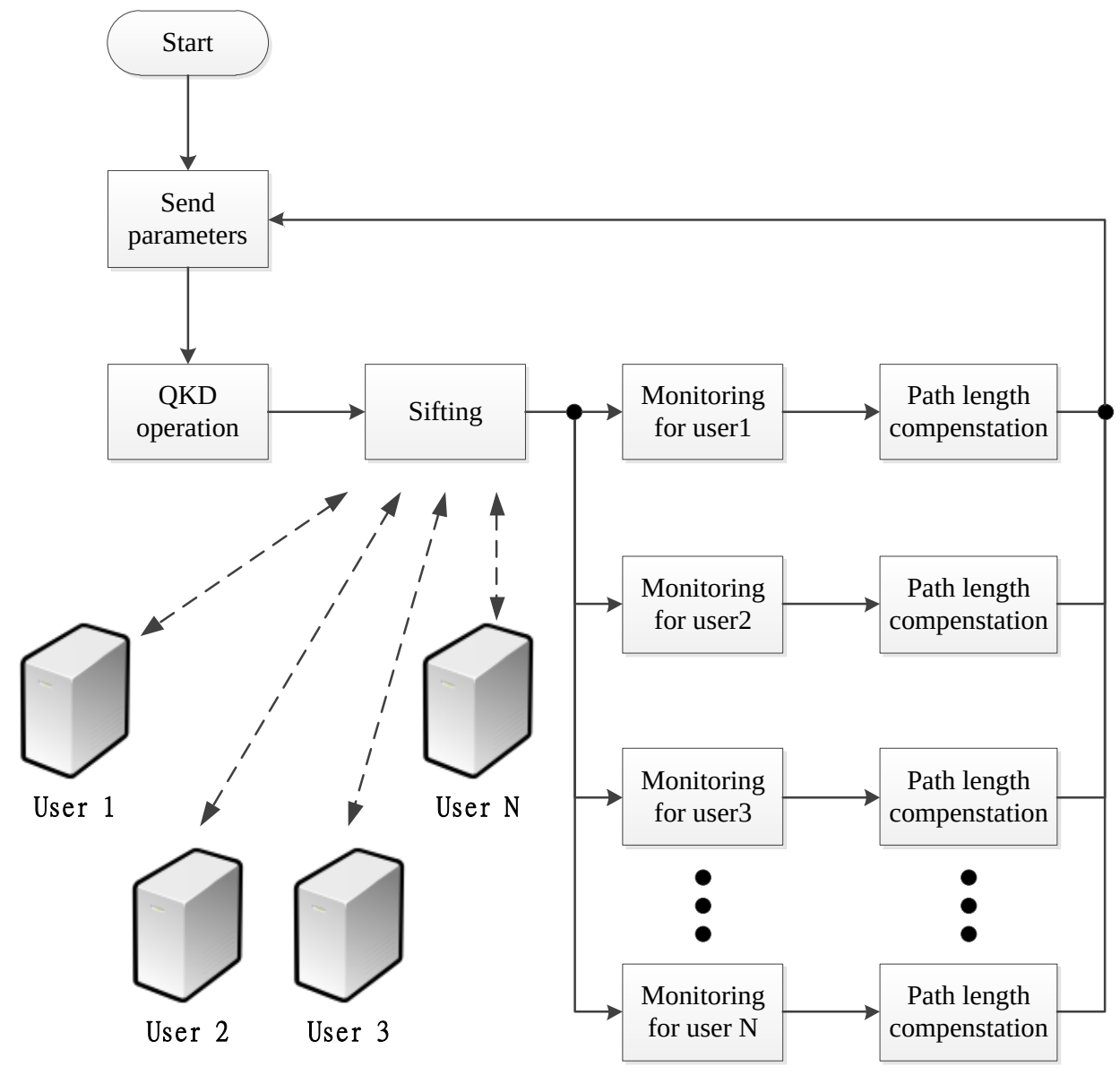

Fig. 3. Flowchart of the control program for laser drivers in the server. After system operation, the control program performs sifting with multiple users. The program corrects the timing parameters using independent monitoring and path length compensation functions.

A control program in the server monitors the system performance parameters, such as the sifted key rate and QBER of the users, to maintain optimal timing of the lasers. The FPGA of the server shares details such as the timing information for the users, digital values to control the PM, and temperatures of the lasers, with the control program. Using this information, the FPGA appropriately operates the QKD network system. The raw key is transferred to the control program and divided for each user by TDM, which will be described later. The server sifts the raw key by communicating with the users, as shown in Fig. 3. After the server and users get the sifted key, the monitoring functions for each user are performed. If the performance is not satisfactory when compared to a threshold level, the path length compensation function is executed using the timing control module. The monitoring and path length compensation functions of the users operate independently through thread and parallel processing, respectively. In the next section, we will discuss the experimental results of the 1 × 64 network system architecture and applications of the timing control modules and control program of the server in a real environment.

## 3. EXPERIMENTS

First, we tested eight of the tunable lasers in the QKD network. As mentioned before, we used WDM and PDM together to enhance user capacity. Half of the lasers have vertical polarization and the other half have horizontal polarization. All lasers with the same polarization have different wavelengths. We measured the wavelengths of the weak coherent lights from the outputs of the two WDM modules, with a spectral resolution of approximately 0.8 nm. Fig. 4 shows the peak wavelengths of the lasers for 64 users. The coherent light signals for 1–32 users have vertical polarization, and their wavelengths are periodically placed from 1532 nm to 1561 nm. The light for 33–64 users has horizontal polarization, and the wavelengths are appropriately allocated as well. Additionally, we checked for crosstalk from other channels. We enabled one channel of the tunable laser and measured the light intensities of the other WDM outputs. All the gaps between the signal and crosstalk noise of the users were at approximately 20 dB, which is very small at the single photon level. Moreover, compared with the pulse width of the laser (2 ns), the time delay due to the dispersion is negligible, and it rarely affects the performance of the system.

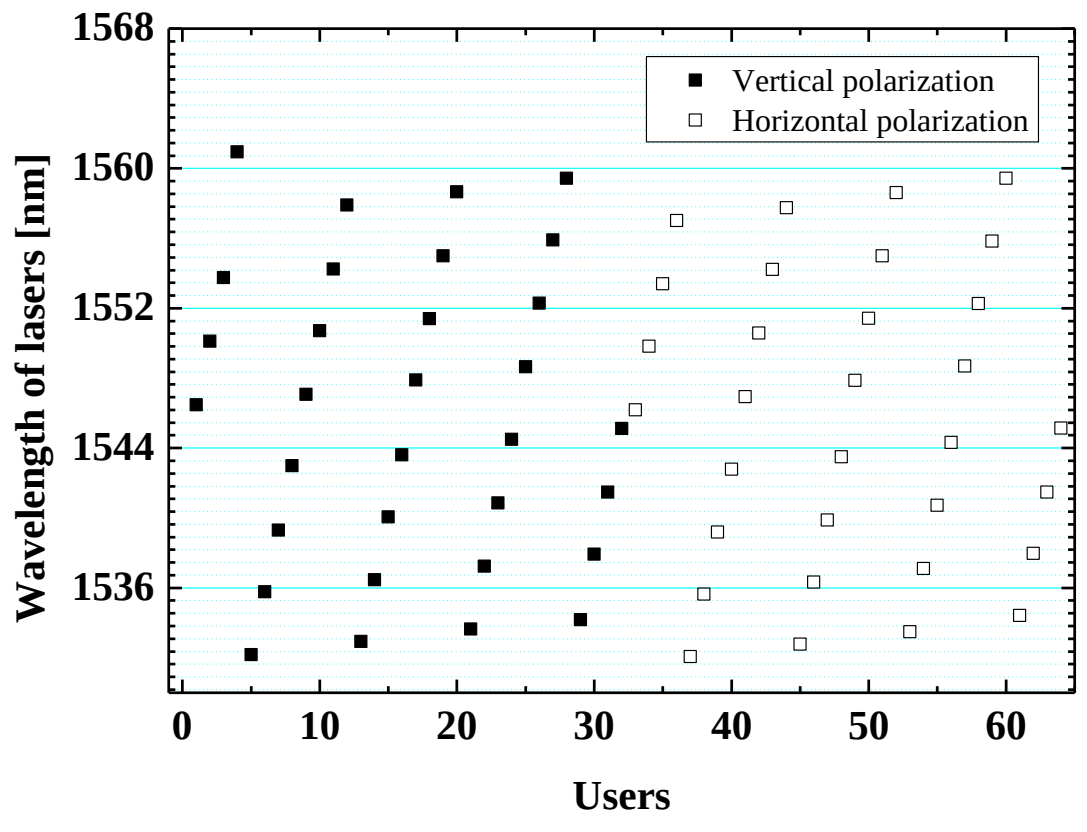

Fig. 4. Peak wavelengths of the lasers for 64 users.

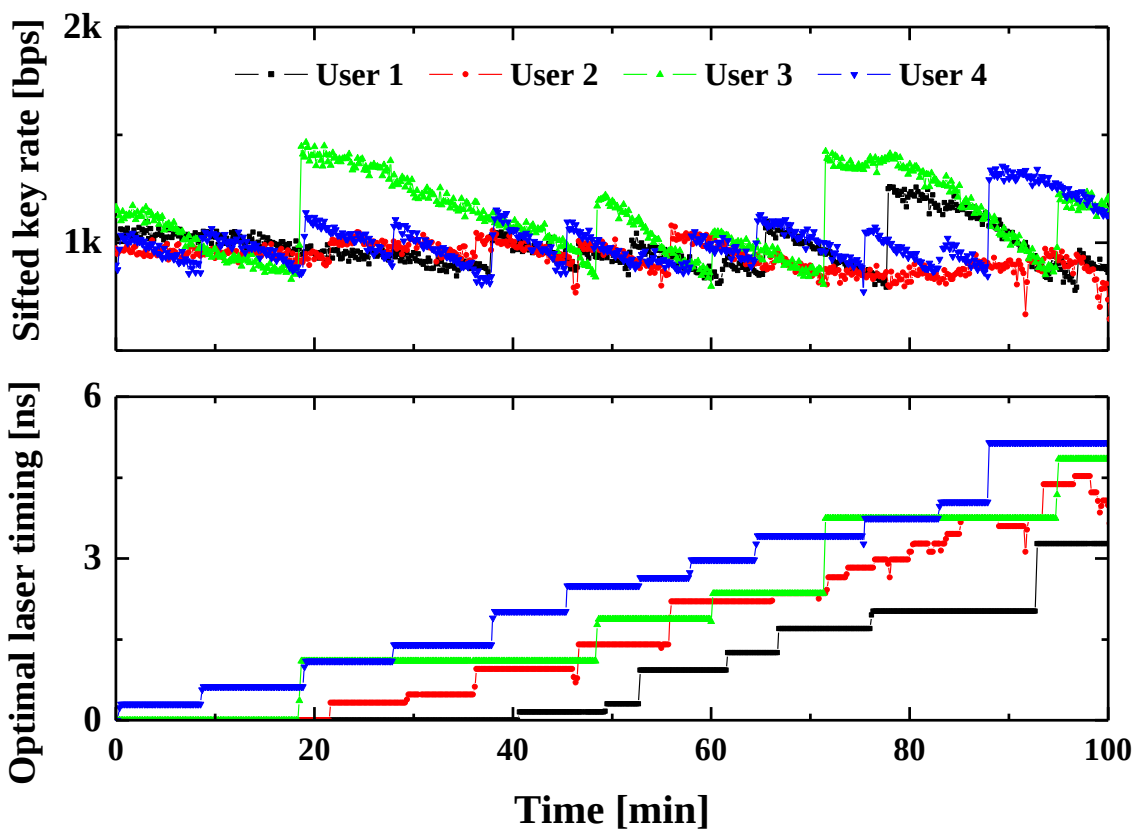

Fig. 5. Sifted key rate and optimal laser timing of four users in 100 min.

To perform experiments using the QKD network system, we developed four transmitters (Alices) in the P&P QKD systems for the users. We verified the functioning of the 1 × 64 QKD network system by changing all WDM outputs with four QKD transmitters in the laboratory. We used four quantum channels with lengths of approximately 25 km but with slight differences. These differences handled by the independent timing control modules of the four users. Using the serializer in the timing control module, we determined the optimal timing of the lasers considering each optical path length of the users. Moreover, owing to temperature changes, the fluctuation in the optimal timing was corrected by delay chains in the timing control module. As shown in Fig. 5, the key rates of each user are independently recovered using the timing control modules and control program over 100 min. Because the flexibility of the optical fiber for each user is different, even in the same location, the operation timings of the path length compensation for different users were not similar. After 100 min, the optimal laser timings for users were changed up to 4–5.2 ns. Therefore, the timing control modules must independently operate to track the optimal laser timing for users in real time. As mentioned above, one of the advantages of the P&P QKD is that server (Bob) knows all the operation timings of the components. Therefore, the laser timings of the users could be appropriately assigned with a 2.5 MHz frequency, considering 10 MHz gate pulses for detection. If one of the users wants to exit the QKD, the other users can increase their key generation speeds by increasing the laser frequency of the remaining users. Moreover, the point-to-point QKD can be operated using this scheme with a maximum laser frequency of 10 MHz for one user. The dark count was under 5000 cps, and the quantum efficiency was 15%. We used 0.6 photons as the signals. The loss of the server is 5.8 dB, including losses of the WDM, PM, interferometer, and circulator. The loss of the QC is approximately 5.2 dB. As all lasers for the users had vertical polarization, we used only one WDM (right-side) and PM ($PM_{B1}$) in this experiment. The PM was randomly modulated using a

(a)

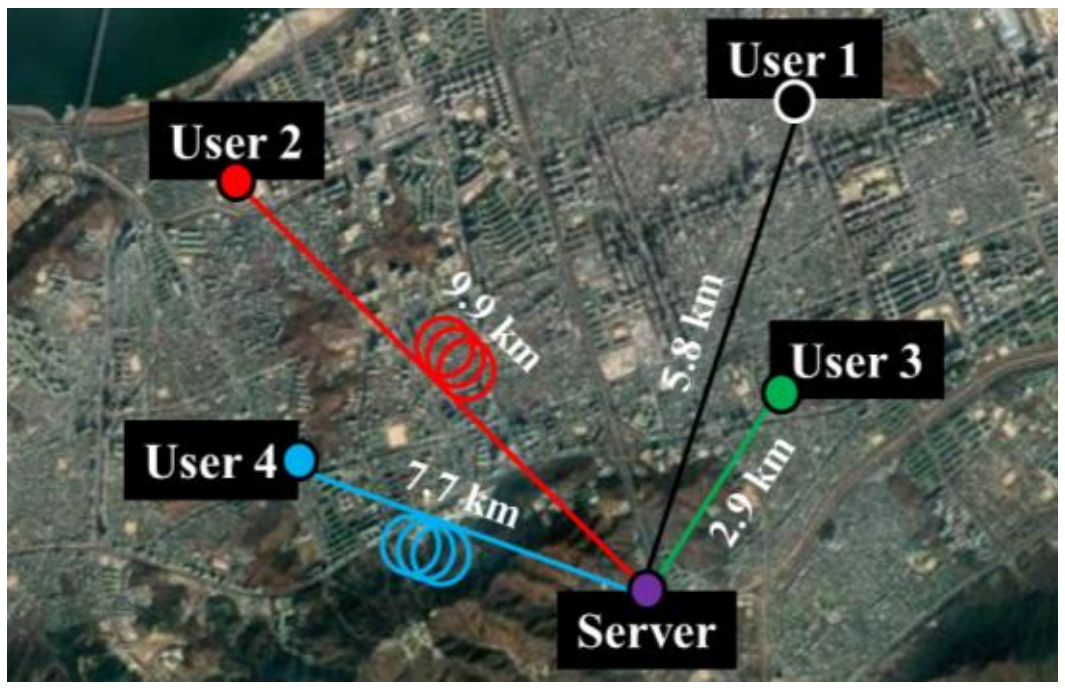

(b)

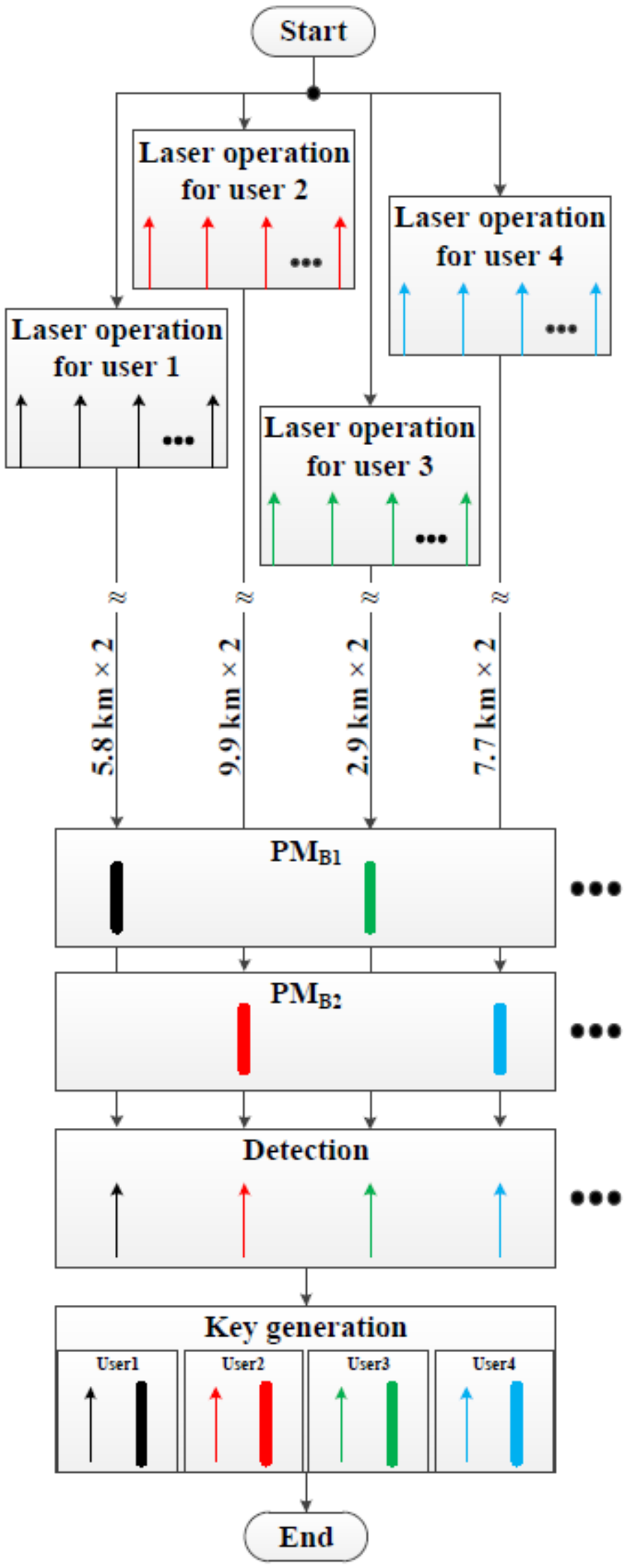

Fig. 6. Map of the 1 × 4 QKD network (a); and operation scheduling for the server to demonstrate the working of the QKD network (b).

quantum random number generator (QRNG) with a 10 MHz modulation frequency.

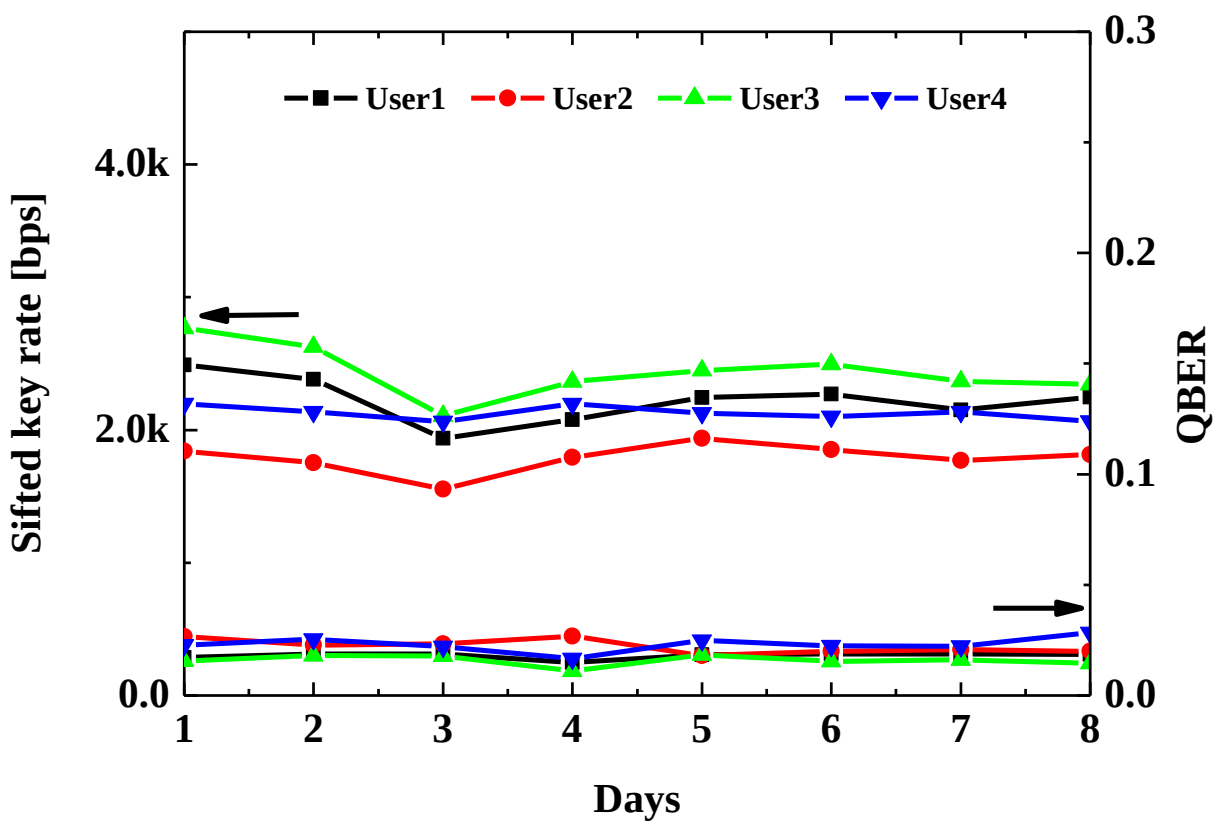

Fig. 7. Results of using the 1 × 4 QKD network over a week in a real-world environment. The points are average key rates and QBERs for one day.

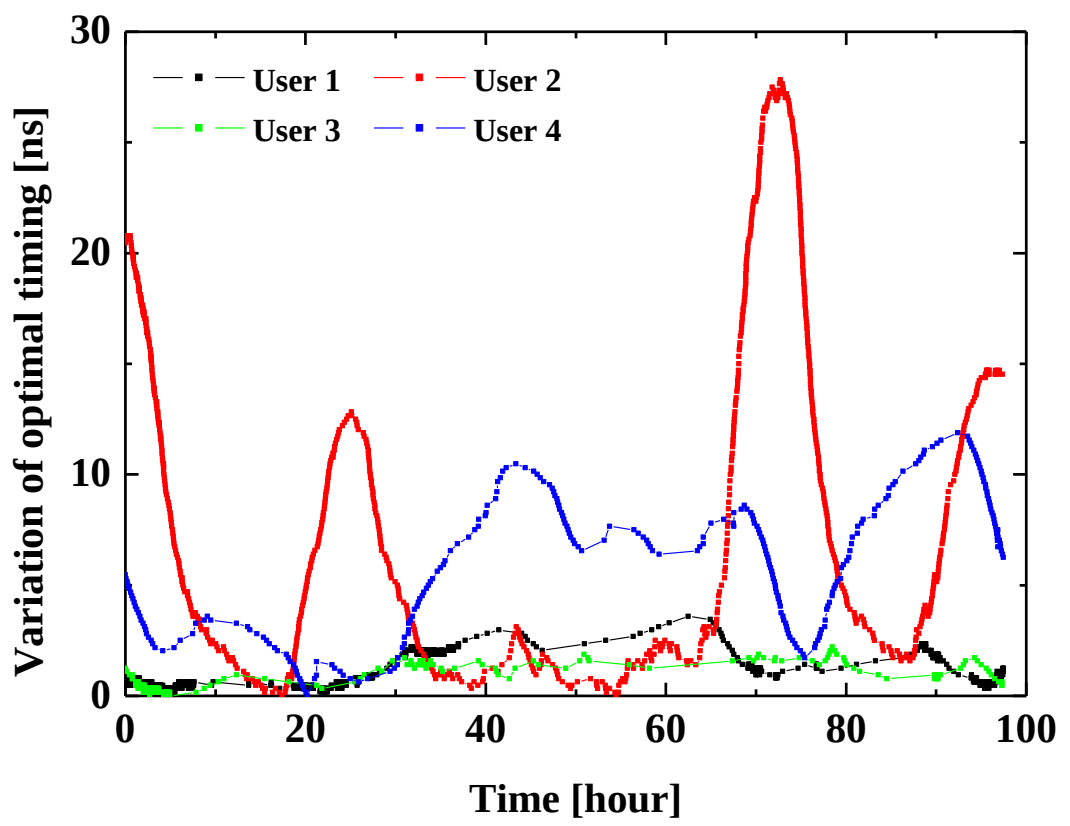

Fig. 8. Variations in optimal laser timings for the real-environment QKD network over 100 h.

Next, we operated the QKD network system in a real-world environment. We placed the server and four users in Seoul city, and the distances between the server and each user are 5.8 km, 9.9 km, 2.9 km, and 7.7 km, as shown Fig. 6 (a). Because the optical path lengths from the server to users are significantly different, we analyzed the system operation timing very carefully. Fig. 6 (b) shows the operation scheduling of the server to distribute quantum keys to the four users in the real-world environment. The lasers can be independently operated by the timing control modules. The laser for each user was operated according to the distance from the server. As user 2 had the longest distance of 9.9 km from the server, the laser for user 2 was operated first. Hence, the sequence of laser operation was 2, 4, 1, 3. Moreover, in this experiment, we used both WDM modules to demonstrate the working of the QKD network using PDM and WDM together. Users 1 and 3 were assigned vertically polarized lasers, and the other users used horizontally polarized lasers. Because the photons from the users pass through the right- or down-side WDM depending on the polarization, the server needs to modulate phases using $PM_{B1}$ (for users 1 and 3) and $PM_{B2}$ (for users 2 and 4) simultaneously, as shown in Fig. 1. The phase modulators were operated at 5 MHz and randomly selected bases using the QRNG. A few nanoseconds after initiating the phase modulation of the photons from the users, the photons arrived at the APDs, operated at 10 MHz of gate pulses. Considering the optical path length for each user, we assigned the photon-arrival timing based on the user numbers at the gate pulse. The first, second, third, and fourth gate pulses indicate the detection timings of the photons from users 1, 2, 3, and 4, respectively. The fifth gate pulse was repeatedly used for detecting the photon from user 1. Therefore, the gate pulse frequency for each user was actually 2.5 MHz; however, for stable operation, we used a fixed gate frequency of 10 MHz. Finally, to generate the sifted key for each user, the raw keys of the server, such as the click information of the APDs and random number for phase modulation, should be divided according to the designated timings among the users. All raw keys were transmitted from the FPGA to the control program and divided for each user. Using the divided raw keys, the server communicated with the users to get the sifted key. Subsequently, with the sifted key rates and QBERs of the users, the control program of the server determined whether the timing of the laser had to be modified to compensate for the change in the optical path length.

We operated the QKD network in a real-world environment over a week, addressing fluctuations in the optical path length using the timing control modules for four users. We set up the QKD network and tested it in a new environment for a week. During this time, we tested all devices under specific experimental conditions. Subsequently, we collected the experimental data. The results of testing the QKD network in a real-world environment are shown in Fig. 7, based on the operation scheduling of the server, as described in Fig. 6 (b). The channel losses for users 1, 2, 3, and 4 were 1.29 dB, 2.23 dB, 1.14 dB, and 1.63 dB, respectively. Apart from conventional channel losses, there were additional channel losses for user 3 because of certain connectors that had to be passed through to reach user 3. However, the other channels had a reasonable loss, lower than 0.23 dB/km. The average sifted key rates for the users were 2.3 kbps, 1.8 kbps, 2.3 kbps, and 2.1 kbps, respectively during a week. In terms of the QBER, all the users had an error rate under 2.5% on average. This shows that the crosstalk due to other channels and environmental noise are negligible; these errors primarily arise from the visibility of interference and detector noise, such as dark counts and after-pulse noise. There are some fluctuations in the key rate, as temperature variations in the server room cause slight differences in the conditions of the APDs and lasers. For stable operation of the QKD network, especially using the WDM modules for a large number of users, the temperature of the lasers is important because of the shifting wavelengths of lasers. We addressed this problem by using feedback control for the temperature of the lasers, as well as by maintaining the temperature of the server room.

We expected the effect of environmental change to be proportional to the optical path length from the server to the user. In Fig. 8, the optimal timing fluctuations of the real-environment QKD network are depicted over 100 h. The optimal timing for user 2, who is farthest from the server, was affected the most by environmental changes, with a maximum adjustment of 28 ns. However, the variation in optimal timing was not exactly proportional to the distance from the server; this is because each user had a storage line that was also affected by the temperature of the user's QKD system. Moreover, there are certain criteria for considering fluctuations in the optimal timing, such as the condition of the lasers, threshold values of the control program, optical losses, and differences in the speed of light caused by the wavelength of the lasers. The optimal timings of users 1 and 3 changed to less than 5 ns, but was approximately 10 ns for user 4. We addressed all the issues encountered when using this system in a real-world environment. The optimal performances of the users were adequately maintained over a week. The optical path length compensation for each user was performed in less than 3 s by using the network architecture based on the P&P system and compact timing control modules and algorithms. Therefore, continuous operation is possible for distributing secure keys to all users.

## 4. CONCLUSION

We developed a 1 × 64 QKD network system with an optical path length compensation and demonstrated its operation, with stable system performance in a real environment. Owing to the effectiveness and stability of the P&P QKD network architecture with its compact control modules based on an FPGA, our system is highly stable for use in one-to-many QKD networks. Moreover, the users do not need expensive components, such as single photon detectors and light sources, to participate in the QKD network. This system can be a favorable candidate for linking end-users or portable devices to organize massive QKD networks.

**Funding Information.** ICT Research and Development Programs of MSIP/IITP (2014-3-00524); National Research Foundation of Korea (NRF) (2019M3E4A107866011, 2019M3E4A1079777, 2019R1A2C2006381); Korea Institute of Science and Technology (KIST) Research Program (2E29580).

**Acknowledgment**. We thank the KT Corporation for providing their deployed fiber.

**Disclosures.** The authors declare no conflicts of interest.

## References

1. C. H. Bennett and G. Brassard, “Quantum cryptography: Public key distribution and coin tossing,” in Proc. IEEE Int. Conf. Comput. Syst. Signal Process. (IEEE, 1984), p. 175–179.
2. A. Ekert, “Quantum cryptography based on Bell’s theorem,” Phys. Rev. Lett. **67**, 661–663 (1991).
3. C. H. Bennett, “Quantum cryptography using any two nonorthogonal states,” Phys. Rev. Lett. **68**, 3121–3124 (1992).
4. A. Muller, H. Zbinden, and N. Gisin, “Quantum cryptography over 23 km in installed under-lake telecom fibre,” Europhys. Lett. **33**, 335–339 (1996).
5. I. Choi, R. J. Young, and P. D. Townsend, “Quantum information to the home,” New J. Phys. **13**, 063039-1–063039-13 (2011).
6. F. Gao, S.J. Qin, W. Huang, and Q.Y. Wen, “Quantum private query: a new kind of practical quantum cryptographic protocols,” Sci. China-Phys. Mech. Astron. **62**, 070301-1–070301-12 (2019).
7. S. Wang, W. Chen, Z. -Q. Yin, Y. Zhang, T. Zhang, H. -W. Li, F.- X. Xu, Z. Zhou, Y. Yang, D. -J. Huang, L. -J. Zhang, F. -Y. Li, D. Liu, Y. -G. Wang, G. -C. Guo, and Z. -F. Han, "Field test of wavelength-saving quantum key distribution network," Opt. Lett. **35**, 2454–2456 (2010).
8. T. -Y. Chen, J. Wang, H. Liang, W. -Y. Liu, Y. Liu, X. Jiang, Y. Wang, X. Wan, W. -Q. Cai, L. Ju, L. -K. Chen, L. -J. Wang, Y. Gao, K. Chen, C. -Z. Peng, Z. -B. Chen, and J. -W. Pan, "Metropolitan all-pass and inter-city quantum communication network," Opt. Express **18**, 27217–27225 (2010).
9. M. Sasaki, M. Fujiwara, H. Ishizuka, W. Klaus, K. Wakui, M. Takeoka, S. Miki, T. Yamashita, Z. Wang, A. Tanaka, K. Yoshino, Y. Nambu, S. Takahashi, A. Tajima, A. Tomita, T. Domeki, T. Hasegawa, Y. Sakai, H. Kobayashi, T. Asai, K. Shimizu, T. Tokura, T. Tsurumaru, M. Matsui, T. Honjo, K. Tamaki, H. Takesue, Y. Tokura, J. F. Dynes, A. R. Dixon, A. W. Sharpe, Z. L. Yuan, A. J. Shields, S. Uchikoga, M. Legré, S. Robyr, P. Trinkler, L. Monat, J.-B. Page, G. Ribordy, A. Poppe, A. Allacher, O. Maurhart, T. Länger, M. Peev, and A. Zeilinger, "Field test of quantum key distribution in the Tokyo QKD Network," Opt. Express **19**, 10387–10409 (2011).
10. S. Wang, W. Chen, Z. -Q. Yin, H. -W. Li, D. -Y. He, Y. -H. Li, Z. Zhou, X. -T. Song, F. -Y. Li, D. Wang, H. Chen, Y. -G. Han, J. -Z. Huang, J. -F. Guo, P. -L. Hao, M. Li, C. -M. Zhang, D. Liu, W. -Y. Liang, C. -H. Miao, P. Wu, G. -C. Guo, and Z. -F. Han, "Field and long-term demonstration of a wide area quantum key distribution network," Opt. Express **22**, 21739–21756 (2014).
11. D. Huang, P. Huang, H. Li, T. Wang, Y. Zhou, and G. Zeng, "Field demonstration of a continuous-variable quantum key distribution network," Opt. Lett. **41**, 3511–3514 (2016).
12. S. -K. Liao, W. -Q. Cai, J. Handsteiner, B. Liu, J. Yin, L. Zhang, D. Rauch, M. Fink, J. -G. Ren, W. -Y. Liu, Y. Li, Q. Shen, Y. Cao, F. -Z. Li, J. -F. Wang, Y. -M. Huang, L. Deng, T. Xi, L. Ma, T. Hu, L. Li, N. -L. Liu, F. Koidl, P. Wang, Y. -A. Chen, X. -B. Wang, M. Steindorfer, G. Kirchner, C. -Y. Lu, R. Shu, R. Ursin, T. Scheidl, C. Z. Peng, J. -Y. Wang, A. Zeilinger, and J. -W. Pan, “Satellite-Relayed Intercontinental Quantum Network,” Phys. Rev. Lett. **120**, 030501-1–030501-4 (2018).
13. Y. -L. Tang, H. -L. Yin, Q. Zhao, H. Liu, X. -X. Sun, M. -Q. Huang, W. -J. Zhang, S. -J. Chen, L. Zhang, L. -X. You, Z. Wang, Y. Liu, C. -Y. Lu, X. Jiang, X. Ma, Q. Zhang, T. -Y. Chen, and J. -W. Pan, “Measurement-device-independent quantum key distribution over untrustful metropolitan network,” Phys. Rev. X **6**, 011024-1–011024-8 (2016).
14. P. Jouguet, S. Kunz-Jacques, T. Debuisschert, S. Fossier, E. Diamanti, R. Allééaume, R. Tualle-Brouri, P. Grangier, A. Leverrier, P. Pache, and P. Painchault, “Field test of classical symmetric encryption with continuous variables quantum key distribution,” Opt. Express **20**, 14030–14041 (2012).
15. K. A. Patel, J. F. Dynes, I. Choi, A. W. Sharpe, A. R. Dixon, Z. L. Yuan, R. V. Penty, and A. J. Shields, “Coexistence of High-Bit-Rate Quantum Key Distribution and Data on Optical Fiber,” Phys. Rev. X **2**, 041010-1–041010-8 (2012).
16. K. A. Patel, J. F. Dynes, M. Lucamarini, I. Choi, A. W. Sharpe, Z. L. Yuan, R. V. Penty, and A. J. Shields, “Quantum key distribution for 10 Gb/s dense wavelength division multiplexing networks,” Appl. Phys. Lett. **104**, 051123-1–051123-4 (2014).
17. P. D. Townsend, S. J. D. Phoenix, K. J. Blow, and S. M. Barnett, “Design of quantum cryptography systems for passive optical networks,” Electron Lett. **30**, 1875–1877 (1994).
18. P. D. Townsend, “Quantum cryptography on multiuser optical fibre networks,” Nature **385**, 47–49 (1997).
19. P. Toliver, R.J. Runser, T.E. Chapuran, J.L. Jackel, T.C. Banwell, M.S. Goodman, R.J. Hughes, C.G. Peterson, D. Derkacs, J.E. Nordholt, L. Mercer, S. McNown, A. Goldman, and J. Blake, "Experimental investigation of quantum key distribution through transparent optical switch elements," IEEE Photon. Technol. Lett. **15**, 1669–1671 (2003).
20. L. Ma, A. Mink, H. Xu, O. Slattery, and X. Tang, “Experimental demonstration of an active quantum key distribution network with over gbps clock synchronization,” IEEE Comm. Lett. **11**, 1019–1021 (2007).
21. G. Brassard, F. Bussieres, N. Godbout, and S. Lacroix, “Multiuser quantum key distribution using wavelength division multiplexing,” in Proc. SPIE (2003), p. 149–153.
22. T. Zhang, X. F. Mo, Z. F. Han, and G. C. Guo, “Extensible router for a quantum key distribution network,” Phys. Lett. A **372**, 3957–3962 (2008).
23. W. Chen, Z. F. Han, T. Zhang, H. Wen, Z. Q. Yin, F. X. Xu, Q. L. Wu, Y. Liu, Y. Zhang, X. F. Mo, Y. Z. Gui, G. Wei, and G. C. Guo, “Field experiment on a star type metropolitan quantum key distribution network,” IEEE Photon. Tech. Lett. **21**, 575–577 (2009).
24. A. Ciurana, J. Martinez-Mateo, M. Peev, A. Poppe, N. Walenta, H. Zbinden, and V. Martin, “Quantum metropolitan optical network based on wavelength division multiplexing,” Opt. Express **22**, 1576–1593 (2014).
25. T. Nishioka, H. Ishizuka, T. Hasegawa, and J. Abe, “Circular type quantum key distribution,” IEEE Photon. Technol. Lett. **14**, 576–578 (2002).
26. P. D. Kumavor, A. C. Beal, S. Yelin, E. Donkor, and B. C. Wang, “Comparison of four multi-user quantum key distribution schemes over passive optical networks,” J. Lightwave Technol. **23**, 268–275 (2005).

27. E. Donkor, “Experimental auto-compensating multi-user quantum key distribution network using a wavelength-addressed bus line architecture,” in Proc. SPIE (2012), p. 839704-1–839704-7.
28. A. Muller, T. Herzog, B. Huttner, W. Tittel, H. Zbinden, and N. Gisin, “Plug and play systems for quantum cryptography,” Appl. Phys. Lett. **70**, 793–795 (1997).
29. G. Ribordy, J. D. Gautier, N. Gisin, O. Guinnard, and H. Zbinden, “Automated ‘plug and play’ quantum key distribution,” Electron. Lett. **34**, 2116–2117 (1998).
30. B. K. Park, M. S. Lee, M. K. Woo, Y. S. Kim, S. W. Han, and S. Moon, “QKD system with fast active optical path length compensation,” Sci. China-Phys. Mech. Astron. **60**, 060311-1–060311-7 (2017).
31. W. -Y. Hwang, “Quantum key distribution with high loss: toward global secure communication,” Phys. Rev. Lett. **91**, 057901-1–057901-4 (2003).
32. X. -B. Wang, “Beating the photon-number-splitting attack in practical quantum cryptography,” Phys. Rev. Lett. **94**, 230503-1–230503-4 (2005).
33. H. -K. Lo, X. Ma, and K. Chen, “Decoy state quantum key distribution,” Phys. Rev. Lett. **94**, 230504-1–230504-4 (2005).
34. X. Ma, B. Qi, Y. Zhao, and H. -K. Lo, “Practical decoy state for quantum key distribution,” Phys. Rev. A **72**, 012326-1–012326-15 (2005).